\documentclass[journal,twoside,twocolumn]{IEEEtran}
\pdfoutput=1  %

\usepackage{amsmath,amssymb}
\usepackage{graphicx}
\usepackage{xcolor}
\usepackage{cite}
\usepackage{balance}
\usepackage{algorithmic}
\usepackage{algorithm}

\usepackage{tikz}
\usepackage{pgfplots}
\pgfplotsset{compat=1.14} %
\usetikzlibrary{arrows,positioning,shadows,shapes,intersections}
\usepackage{hyperref}
\hypersetup{
  breaklinks=true,  %
  colorlinks=true,  %
  citecolor=blue,  %
  linkcolor=blue  %
}

\renewcommand\arraystretch{1.3}  %

\newcommand{\E}{{\mathbb{E}}}
\newcommand{\R}{{\mathbb{R}}}
\newcommand{\Z}{{\mathbb{Z}}}
\newcommand{\T}{{\mathrm{T}}}

\newcommand{\bA}{{\boldsymbol{A}}}
\newcommand{\tA}{\boldsymbol{\tilde{A}}}
\newcommand{\bB}{{\boldsymbol{B}}}
\newcommand{\bC}{{\boldsymbol{C}}}
\newcommand{\be}{{\boldsymbol{e}}}
\newcommand{\bE}{{\boldsymbol{E}}}
\newcommand{\bI}{{\boldsymbol{I}}}

\newcommand{\bR}{{\boldsymbol{R}}}
\newcommand{\bu}{{\boldsymbol{u}}}
\newcommand{\bU}{{\boldsymbol{U}}}
\newcommand{\bx}{{\boldsymbol{x}}}
\newcommand{\bX}{{\boldsymbol{X}}}
\newcommand{\by}{{\boldsymbol{y}}}
\newcommand{\bz}{{\boldsymbol{z}}}
\newcommand{\bzero}{{\boldsymbol{0}}}
\newcommand{\dx}{\,\mathrm{d}\bx}
\newcommand{\uhat}{{\hat{\bu}}}
\newcommand{\cL}{{\mathcal{L}}}
\newcommand{\bareps}{{\bar\epsilon}}
\newcommand{\h}{\tfrac{1}{2}} %

\newcommand{\tr}{T_\mathrm{r}}
\newcommand{\uran}{\mathit{URAN}}
\newcommand{\gran}{\mathit{GRAN}}
\newcommand{\clp}{\mathit{CLP}}
\newcommand{\red}{\mathit{RED}}
\newcommand{\orth}{\mathit{ORTH}}

\DeclareMathOperator*{\argmin}{arg\,min} %
\DeclareMathOperator{\sgn}{sgn}
\DeclareMathOperator{\adj}{adj}

\newcommand{\eqlab}[2]{\begin{align} \label{#1} #2 \end{align}}
\newcommand{\eq}[1]{\begin{align} #1 \end{align}}

\newtheorem{example}{Example}

\title{Optimization and Identification \\ of Lattice Quantizers}
\author{Erik Agrell, \IEEEmembership{Fellow, IEEE}, Daniel Pook-Kolb, and Bruce Allen, \IEEEmembership{Member, IEEE}
\thanks{The work of E.~Agrell was supported by a Collaborating Scientist Grant from the Max Planck Institute for Gravitational Physics, Germany, which is gratefully acknowledged.}%
\thanks{E.~Agrell is with the Department of Electrical Engineering, Chalmers University of Technology, 41296 Gothenburg, Sweden (e-mail: agrell@chalmers.se).}
\thanks{D.~Pook-Kolb and B.~Allen are with the Max Planck Institute for Gravitational Physics, 30167 Hannover, Germany, and Leibniz Universit\"at Hannover (e-mail: daniel.pook.kolb@aei.mpg.de and bruce.allen@aei.mpg.de).}
}

\begin{document}

\maketitle

\begin{abstract} %
Lattices with minimal normalized second moments are designed using a new numerical optimization algorithm. Starting from a random lower-triangular generator matrix and applying stochastic gradient descent, all elements are updated towards the negative gradient, which makes it the most efficient algorithm proposed so far for this purpose. A graphical illustration of the theta series, called theta image, is introduced and shown to be a powerful tool for converting numerical lattice representations into their underlying exact forms. As a proof of concept, optimized lattices are designed in dimensions up to 16. In all dimensions, the algorithm converges to either the previously best known lattice or a better one. The dual of the 15-dimensional laminated lattice is conjectured to be optimal in its dimension and its exact normalized second moment is computed.
\end{abstract}


\section{Introduction} \label{s:intro}
\IEEEPARstart{A}{classical} problem in geometry is how to construct a structure of infinitely many points in $n$-dimensional space so that the mean square distance between a random real vector and its closest member of the point structure is minimal, for a fixed density of points per unit volume. Apart from its fundamental theoretical value, the problem has important practical applications in, e.g., digital communications (both source \cite{gray98} and channel coding \cite[p.~70]{conway99splag}, \cite[Ch.~9]{zamir14book}), pattern recognition \cite{tuytelaars07}, cryptography \cite{guo15}, hashing \cite{jegou08}, machine learning \cite{sablayrolles19}, and data analysis \cite{allen21prd}.

The construction of structures with small mean square distance is traditionally called the \emph{quantizer problem} and the figure of merit is the (possibly normalized) \emph{second moment.} The problem was first formulated in 1959 by Fejes T\'oth, who also solved it for the two-dimensional case \cite{fejestoth59}. The optimal structure in two dimensions is, not surprisingly, the hexagonal lattice.
Gersho in 1979 suggested lattice structures in three and four dimensions \cite{gersho79}, which still today offer the smallest known normalized second moments (NSM) known in their respective dimensions. The proposed three-dimensional structure, which is the body-centered cubic lattice, was proved to be optimal among three-dimensional lattices by Barnes and Sloane \cite{barnes83}. Lattices with small NSM are called \emph{lattice quantizers.}

In 1982, Conway and Sloane calculated the NSMs of the most common lattice families and tabulated the best known lattice quantizers in dimensions up to $10$ \cite{conway82voronoi}. They further conjectured that the best lattice quantizer in any dimension is the dual of the lattice that solves the more well-studied \emph{packing problem.} Two years later, the same authors found improved lattices in dimensions $6$ and $7$ and numerically computed the NSMs of known lattices in dimensions $12$, $16$, and $24$, which all supported their duality conjecture \cite{conway84}. The pioneering lattice research by Conway and Sloane is summarized in their now classic textbook \cite{conway99splag}.

Improved lattice quantizers in dimensions $9$ and $10$ (as well as improved nonlattice structures in dimensions $7$ and $9$) were found by Agrell and Eriksson in 1998 using a numerical optimization technique \cite{agrell98}. These lattices are not duals of the densest lattice packings known, and thus indicate that Conway and Sloane's conjecture might not be true. Dutour Sikiri\'c \emph{et al.} found an improved lattice in $11$ dimensions and evaluated its NSM exactly, thus providing further evidence against the conjecture \cite{dutour09}. Allen and Agrell described exactly the conjectured optimal $9$-dimensional lattice quantizer, which was previously only approximated numerically \cite{allen21adp}. Recently, some improved lattices in dimensions $12$ and above have been reported \cite{lyu22, agrell23, agrell24, pook-kolb24arxiv}.

While every lattice for which the exact NSM has been computed serves to constructively prove the achievability of that particular NSM value, there is no known ``converse theorem'' in the sense of proving that no better lattice exists. Lower bounds on the NSM are known or conjectured \cite{conway85}, but they are not tight in any dimension above $2$.

In this paper, we take a fresh look at numerical optimization of lattice quantizers, which as far as we know has not seen any progress since 1998 \cite{agrell98}. A more efficient algorithm for the purpose is proposed. Like the previous algorithm, the new one employs stochastic gradient descent (SGD), which is a standard technique in numerical optimization. The main difference lies in how the gradient is computed. In the previous algorithm, one degree of freedom is reserved to maintain a fixed number of lattice points per unit volume, while the other degrees of freedom are updated guided by the gradient of the \emph{unnormalized} second moment. Thereby the NSM decreases, but unfortunately not in the direction of its gradient. This shortcoming is resolved in the new algorithm, which makes it more efficient.

We also address the question of identifying an exact form for a lattice that is only known numerically. To this end, we introduce a new graphical tool, called \emph{theta image,} and use it to derive a linear system of equations whose solution gives the desired exact form in many cases. Even in cases where the method does not yield an exact form, most of the unknown degrees of freedom are identified as functions of a small number of unknown parameters. 

As a proof of concept, lattices with conjectured minimal NSMs are constructed in dimensions up to $16$, and exact forms are identified for all of them. The results revisit previously best known lattice quantizers in most considered dimensions and reveals a new best lattice quantizer in $15$ dimensions.

The paper is organized as follows. Sec.~\ref{s:lattices} compactly summarizes relevant lattice definitions and theory. The new algorithm is theoretically developed in Sec.~\ref{s:sgd}. Sec.~\ref{s:implementation} provides pseudocode for the algorithm along with some implementational considerations. This section is mainly aimed at programmers and more or less stand-alone from the theoretical foundation in Sec.~\ref{s:sgd}. Other possible algorithms for numerical optimization of lattice quantizers are reviewed in Sec.~\ref{s:algorithms}, where also the improvements over \cite{agrell98} are discussed. In Sec.~\ref{s:identification}, we discuss how the exact form of a lattice that is only known approximatively can be obtained using the theta series. Sec.~\ref{s:poc}, finally, demonstrates the potential of the new framework by constructing conjectured optimal lattice quantizers.

\section{Lattices and Their Second Moments} \label{s:lattices}

\emph{Notation:}
The norm (Euclidean length) of a vector $\bx$ is denoted by $\|\bx\|$. The $n\times n$ identity matrix is denoted by $\bI_n$. The $i$th element of a vector $\bx$ is denoted by $[\bx]_i$, while the element in row $i$, column $j$ of a matrix $\bX$ is denoted by $[\bX]_{i,j}$. All vectors are row vectors.

Given a set of $n$ linearly independent \emph{basis vectors} in $\R^m$, a \emph{lattice} $\cL$ is the set of all linear combinations of these vectors with integer coefficients. The basis vectors are collected as rows in an $n\times m$ \emph{generator matrix} $\bB$, where the condition of linearly independent rows implies $m\ge n$. Expressed in terms of the generator matrix, the lattice $\cL$ consists of the points $\bu\bB$ for all $\bu \in \Z^n$. By definition, the all-zero vector $\bzero$ belongs to any lattice. The inner products of all basis vectors with each other are collected in the symmetric, positive definite \emph{Gram matrix} $\bA=\bB \bB^\T$.

The basis vectors span an $n$-dimensional Euclidean space, which is a linear subspace of $\R^m$. 
The set of points in this subspace that are closer to $\bzero$ than to any other point in $\cL$ is the \emph{Voronoi region} $\Omega$ of the lattice. The Voronoi region $\Omega$ is a \emph{fundamental cell} of $\cL$, which means that infinitely many copies of $\Omega$, each translated by a vector in $\cL$, fill the $n$-dimensional space in which the lattice resides without gaps or overlaps, except on the boundaries. Another fundamental cell of $\cL$ is the parallelotope $[0,1)^n \bB$.

The standard figures-of-merit for lattices can all be expressed in terms of properties of $\Omega$. In this paper, we focus on the NSM \cite{gersho79}, \cite[pp.~34, 56--62]{conway99splag}
\eqlab{e:gdef}{
G = \frac{1}{nV^{1+2/n}} \int_\Omega \|\bx\|^2 \dx,
}
where $V = (\det\bA)^{1/2}$ is the $n$-volume of $\Omega$. The normalization coefficient $1/V^{1+2/n}$ ensures that $G$ is dimensionless and hence invariant to lattice scaling, while the coefficient $1/n$ ensures that lattices constructed as the product of identical lower-dimensional lattices have the same NSM as these constituent lattices \cite[Sec.~IV]{agrell23}. For example, the cubic lattice $\Z^n$ has $G=1/12$ for any $n$.

Two lattices are \emph{equivalent} in Euclidean space if their Voronoi regions are congruent, and two generator matrices are equivalent if they generate equivalent lattices. Equivalent lattices have the same NSM. The packing density, covering density, kissing number, and all other metrics invariant to scaling and orthogonal transformation are also the same. Mathematically, two generator matrices $\bB$ and $\bB'$ of sizes $n\times m_1$ and $n \times m_2$, respectively, where $m_1 \le m_2$, are equivalent if and only if \cite[p.~10]{conway99splag}
\eqlab{e:equiv}{
\bB' = c \, \bU \bB \bR
,}
where $c$ is a real scalar, $\bU$ is an $n\times n$ integer matrix with determinant $\pm 1$, and $\bR$ is a real $m_1 \times m_2$ matrix with orthonormal rows, i.e., $\bR\bR^\T = \bI_{m_1}$.
Left-multiplication by $\bU$ corresponds to a basis change and right-multiplication by $\bR$ to an orthogonal transformation, i.e., rotation and/or reflection.
The corresponding relation between Gram matrices is
\eqlab{e:equivg}{
\bA'= c^2 \bU \bA \bU^\T
.}

For every generator matrix $\bB'$ of size $n\times m_2$, there exists a square generator matrix $\bB$ that satisfies \eqref{e:equiv} with $m_1=n$. The analysis of lattices in Euclidean space can therefore without loss of generality be confined to square generator matrices, which is exploited in this work.

Another important characteristic of a lattice $\cL$ is the \emph{theta series} \cite[pp.~44--47]{conway99splag}
\eqlab{e:theta}{
  \theta(q) = \sum_{\bx\in\cL} q^{\|\bx\|^2} = 1+\tau q^{4\rho^2}+\cdots
,}
where $q$ is a dummy variable. It characterizes how the lattice points
are distributed on spherical \emph{shells} around the origin. The
first term arises from the lattice vector $\bzero$.  The second term
contains information about both the \emph{kissing number} $\tau$,
i.e., the number of shortest nonzero lattice vectors, and the
\emph{packing radius} $\rho$, i.e., half the minimal distance between
two lattice vectors. Equivalent lattices have the same theta series.

\section{Stochastic Gradient Descent} \label{s:sgd}

The starting point for minimization by stochastic gradient descent (SGD) is to write the objective function as the expected value of a real function of (i) a set of random variables with a given distribution and (ii) a set of deterministic variables, whose values are to be optimized. The gradient of this real function with respect to the deterministic variables is calculated, or if the exact gradient is not known, the gradient of a Monte-Carlo estimate of the objective function serves as a surrogate. The distribution of the random variables does not have to be known analytically, as long as an algorithmic procedure is available to generate random samples. Independent samples are generated sequentially using this procedure, and for every sample, the deterministic variables are shifted a small step in the direction of the negative gradient. In this section, theory is developed for applying SGD to NSM minimization. This leads to our lattice construction algorithm, whose implementation is described in the next section.

\subsection{Second Moment Estimation} \label{s:estimation}

An elegant method to generate random vectors uniformly in the Voronoi region $\Omega$ of a given lattice $\cL$ was proposed in \cite{conway84} for the purpose of NSM estimation. Let $\bz$ be a random vector drawn uniformly from the unit $n$-cube $[0,1)^n$ and let, for a given generator matrix $\bB$,
\eqlab{e:uhat}{
\uhat = \argmin_{\bu\in\Z^n} \|(\bz-\bu)\bB\|^2.
}
Now $\uhat\bB$ is the lattice point closest to $\bz\bB$ (which is normally not a lattice point). Since the Voronoi region $\Omega$ and the parallelotope $[0,1)^n \bB$ are both fundamental cells of $\cL$, $\be=(\bz-\uhat)\bB$ is uniformly distributed in $\Omega$. The generation of uniform random vectors in $\Omega$ is discussed in detail in \cite[Sec.~4.2]{zamir14book}.

To calculate \eqref{e:uhat} requires solving the \emph{closest point problem} for a given lattice. Algorithms for this purpose are available for classical, well-structured lattices \cite{conway82decoding, conway84} as well as arbitrary lattices \cite{agrell02, ghasemmehdi11}.

Using these definitions of $\bz$, $\uhat$, and $\be$, the NSM in \eqref{e:gdef} can be written as
\eqlab{e:gexp}{
G = \E_\bz[g(\bB,\bz)],
}
where
\eqlab{e:gz}{
g(\bB,\bz) = \frac{1}{n}V^{-2/n} \|\be\|^2.
}
Here $V$ is a function of $\bB$ and $\be$ is a function of both $\bB$ and $\bz$.
Note that even if \eqref{e:uhat} has multiple solutions, $g(\bB,\bz)$ is still well-defined,
because $\|\be\|^2$ has the same value for all solutions $\uhat$.  These multiple solutions
occur iff  $\bz \bB$ lies on the boundary of $\Omega$ (or its images
under translation by the lattice).

If $\bz_1,\ldots,\bz_T$ denote $T$ independent realizations of $\bz$, then an unbiased estimate of $G$ follows immediately from \eqref{e:gexp} as
\eqlab{e:ghat}{
\hat{G} = \frac{1}{T}\sum_{t=1}^T g(\bB,\bz_t)
.}
To quantify the estimation accuracy, we use an unbiased estimator of the variance of $\hat{G}$ \cite[Sec.~IV]{pook-kolb23}
\eqlab{e:sighat}{
{\hat\sigma}^2 = \frac{1}{T-1} \left(\frac{1}{T}\sum_{t=1}^T g^2(\bB,\bz_t)-\hat{G}^2\right) \,,
}
which is much more accurate than the ``jackknife'' estimator recommended in earlier literature.

It is easily verified that \eqref{e:ghat} remains unchanged if the lattice, represented by $\bB$, is rescaled. However, previous descriptions of the same NSM estimation method are valid only for lattices with $V=1$. This is because of an unfortunate error in the original publication \cite{conway84}, where the right-hand sides of \cite[Eqs.~(2), (4)]{conway84} are missing a factor corresponding to the volume of the Voronoi region (here denoted by $V$). This error appears to have propagated to \cite[Eqs. (73)--(74)]{lyu22} and \cite[Eqs.~(12)--(15)]{pook-kolb23}.

\subsection{Iterative basis vector update} \label{s:update}

If the gradient of the NSM $\nabla_{\bB}G$ were known analytically for arbitrary $\bB$, then this gradient could be used to minimize $G$. In a few special cases where we know $G$ exactly, it is an infinitely differentiable ($C^\infty$) function of certain elements of $\bB$ in a certain domain, or \emph{phase,} within which the overall topology of $\Omega$ (by which we mean the number of faces in any dimension and their pattern of intersection) is fixed. Furthermore, it is many times differentiable (approximately $n$ times) on the boundaries between such domains, called \emph{phase transitions} \cite{allen21adp, pook-kolb24arxiv}. If these examples are indicative of how $G$ behaves in general, then the function $G(\bB)\colon\R^{n\times m} \to \R$ is $C^\infty$ except on these phase transitions. Thus,
generic values of $\bB$ have open neighborhoods in which $G$ is smooth, so gradient descent could be effective. We are however not aware of any general expression for $\nabla_{\bB}G$.\footnote{
\emph{Note added in proof:} 
It is shown in \cite[Th.~3.1]{regev22arxiv} that the gradient of the unnormalized second moment is twice the covariance matrix \eqref{e:covariance}, demonstrating that $G(\bB)$ and its gradient are continuous. However, since estimating the covariance matrix is computationally costly (see Sec.~\ref{s:whitening}), this is not immediately useful for gradient-based NSM optimization.}

In the absence of the exact $\nabla_{\bB}G$, the NSM estimate $\hat{G}$ in \eqref{e:ghat} can be straightforwardly differentiated and is well suited for minimization by SGD.
Specifically, we generate a sequence of random vectors $\bz$, and for each $\bz$, we calculate $\uhat$ and $\be$. Then the generator matrix $\bB$ is updated a small step in the direction that decreases $g(\bB,\bz)$ most, which is the negative gradient with respect to the elements of $\bB$.

By a suitable choice of coordinate system, every lattice can be represented by a \emph{square, lower-triangular} generator matrix with \emph{positive diagonal elements.} Therefore, without loss of generality, our lattice construction method considers only such generator matrices. From now on and until the end of Sec.~\ref{s:implementation}, the generator matrix $\bB$ has size $n\times n$ and its elements satisfy $[\bB]_{i,j} = 0$ for $i<j$ and $[\bB]_{i,j} > 0$ for $i=j$. Apart from reducing the number of degrees of freedom from $nm$ to $n(n+1)/2$, these conditions greatly facilitate the calculation of the volume $V$ as a function of $\bB$, which is now simply
\eqlab{e:vol}{
V = \prod_{k=1}^n [\bB]_{k,k}.
}
Furthermore, the closest point algorithms we use to compute $\be$ from $\bz$ need a triangular matrix $\bB$ \cite{agrell02, ghasemmehdi11}.

We are now ready to calculate the gradient of $g(\bB,\bz)$
in \eqref{e:gz}, which consists of the partial derivatives
\eqlab{e:dgz}{
\frac{\partial g(\bB,\bz)}{\partial [\bB]_{i,j}} =
	\frac{1}{n} V^{-2/n} \frac{\partial \|\be\|^2}{\partial [\bB]_{i,j}}
	+ \frac{1}{n} \|\be\|^2 \frac{\partial V^{-2/n}}{\partial [\bB]_{i,j}}
        .}
Note that for the multiple solution case described after \eqref{e:gz},
the first partial derivative on the right-hand side of \eqref{e:dgz} is
undefined. For generic $\bB$, when $\bz \bB$ crosses the boundary of
$\Omega$, the derivative steps abruptly from a well-defined value on
one side of the boundary to a different well-defined value on the
other side.  For a given $\bB$, the set of $\bz$ for which this occurs
is a set of measure zero.  Thus, from here forward, we assume that
$\bz$ denotes one of the $T$ values $\bz_1,\ldots\bz_T$, which
do not lie in this set of measure zero.

We define $\by=\bz-\uhat$, so that $\be = \by\bB$. For a given $\bz$ and infinitesimal changes of $\bB$, $\uhat$ and $\by$ are unchanged.
Hence $\partial \by/\partial [\bB]_{i,j} = -\partial \uhat/\partial [\bB]_{i,j}=0$ for all $i$ and $j$. Therefore,
\eq{
\frac{\partial \|\be\|^2}{\partial [\bB]_{i,j}} &=
	2 \frac{\partial \be}{\partial [\bB]_{i,j}} \be^\T \notag\\
	&= 2[\by]_i [\be]_j, \label{e:de} \quad \textrm{if $i \ge j$}, \\
\frac{\partial V^{-2/n}}{\partial [\bB]_{i,j}} &=
	\frac{\partial}{\partial [\bB]_{i,j}} \prod_{k=1}^n [\bB]_{k,k} \notag\\
	&= 0, \quad \textrm{if $i>j$}. \label{e:dvii}
}
For the diagonal elements,
\eqlab{e:dvij}{
\frac{\partial V^{-2/n}}{\partial [\bB]_{i,i}} &=
	\frac{\partial}{\partial [\bB]_{i,i}} \prod_{k=1}^n [\bB]_{k,k}^{-2/n} \notag\\
	&= -\frac{2}{n} [\bB]_{i,i}^{-1-2/n} \prod_{k\ne i} [\bB]_{k,k}^{-2/n} \notag\\
	&= -\frac{2V^{-2/n}}{n[\bB]_{i,i}}.
}
Combining \eqref{e:dgz}, \eqref{e:de}, \eqref{e:dvii}, and \eqref{e:dvij} yields
\eqlab{e:dgz2}{
\frac{\partial g(\bB,\bz)}{\partial [\bB]_{i,j}} = \begin{cases}
	\frac{2}{n} V^{-2/n} [\by]_i [\be]_j, & i>j, \\
	\frac{2}{n} V^{-2/n} \left( [\by]_i [\be]_i-\frac{\|\be\|^2}{n[\bB]_{i,i}} \right), & i=j.
\end{cases}
}

Now introducing explicit notation for the steps in the SGD algorithm, we denote the $T$ random realizations of $\bz$ with $\bz_t$ for $t=0,\ldots,T-1$, with $t$ being referred to as ``time.'' In each time instant $t$, the generator matrix $\bB = \bB_t$ is updated as
\eqlab{e:update}{
[\bB_{t+1}]_{i,j} = [\bB_t]_{i,j} - \epsilon \frac{\partial g(\bB_t,\bz_t)}{\partial [\bB_t]_{i,j}}
}
for $i,j = 1, \ldots, n$ and $i \ge j$, where $\epsilon$ is a small step size.\footnote{With time-dependent $\bz$ and $\bB$, other quantities such as $V$ and $\be$ also become time-dependent, although we will still write them without explicit subscripts $t$.}

As usual when gradient descent is applied to a smooth function, the iterations converge for small enough step sizes. If a too large step is taken in \eqref{e:update}, then $\bB$ may move to the other side of the ``valley'', where $g(\bB,\bz)$ increases again. For very large steps, a diagonal element $[\bB]_{i,i}$ may even become negative, in which case the optimization fails. To avoid this, $\epsilon$ should be chosen small enough. What is small enough depends greatly on the structure of the generator matrix $\bB$. Generally, an ill-conditioned $\bB$ is more sensitive to the size of $\epsilon$, whereas a \emph{reduced} $\bB$ is more stable. A reduced generator matrix is one whose rows (basis vectors) are relatively short and orthogonal to each other, according to some heuristic criterion \cite{lenstra82}. We therefore apply \emph{lattice reduction} to $\bB$ regularly throughout the iterations \cite{agrell98}. Reduction also brings the added benefit of speeding up the closest point search algorithm \cite{agrell02}, which dominates the complexity of the optimization procedure. On the other hand, reduction generally brings the generator matrix away from the desired lower-triangular form, but this can be alleviated by rotating the coordinate system.

The elements of the initial generator matrix $\bB_0$ are chosen as independent zero-mean unit-variance Gaussian random variables, immediately followed by reduction and orthogonal transformation to bring it into the desired lower-triangular form with positive diagonal elements. Repeating the optimization procedure multiple times with different $\bB_0$ enables the algorithm to find local minima in different parts of the optimization space, which is essential when searching for a global minimum of the nonconvex function $G$. There are infinitely many global minima in any dimensions $n$, reflecting the fact that equivalent generator matrices \eqref{e:equiv} yield the same NSM $G$.

\subsection{Step size} \label{s:stepsize}

The convergence of the algorithm depends strongly on the choice of step size $\epsilon$ in \eqref{e:update}. The step size may be selected as a function of $\bB_t$ and $t$ (but not $\bz_t$, $i$, or $j$). In our optimization algorithm, we define
\eqlab{e:eps}{
  \epsilon = (n/2)V^{2/n} \mu_t
,}
where $\mu_t$ is an annealing scheme to be discussed later.

The factor $V^{2/n}$ in \eqref{e:eps} makes the algorithm insensitive to the scale of $\bB$. To see this, consider how the update rule \eqref{e:update} affects a family of generator matrices $\bB_t = \bB = c\bar\bB$ for any fixed generator matrix $\bar\bB$ and a variable scale factor $c>0$.
For a given $\bz$, $\be$ scales linearly with $c$, $V$ scales with $c^n$, and $g(\bB,\bz)$ in \eqref{e:gz} remains constant. Furthermore, the gradient \eqref{e:dgz2} scales with $c^{-1}$ and $\epsilon$ in \eqref{e:eps} scales with $c^2$. Therefore, $\bB_{t+1}$ in \eqref{e:update} scales linearly with $c$ if $\bB_t$ scales with $c$, as desired. The coefficient $n/2$ is not important and only serves to simplify the notation in Sec.~\ref{s:implementation}.

Substituting \eqref{e:eps} into \eqref{e:update}, the update rule becomes
\eqlab{e:muupdate}{
[\bB_{t+1}]_{i,j} = [\bB_t]_{i,j} - \mu_t \frac{n V^{2/n}}{2} \frac{\partial g(\bB_t,\bz_t)}{\partial [\bB_t]_{i,j}}
,}
where the coefficient $(n/2)V^{2/n}$ is conveniently absorbed by its inverse in \eqref{e:dgz2}.

The purpose of $\mu_t$ is that the speed of movement in the variable space should decrease gradually with time, which is why it is called an annealing (or cooling) scheme. A large step size is beneficial initially, allowing $\bB_t$ to move fast to the vicinity of a minimum and escape possible plateaus and shallow local minima, whereas a small step size is better near the end of the optimization process, in order to fine-tune $\bB_t$ towards the exact minimum.

As usual in SGD optimization, there is no guarantee that the algorithm will converge to the global optimum, or even to a local minimum. It may with a nonzero probability terminate literally anywhere in the vast space of generator matrices $\bB$. The annealing scheme and other optimization parameters should be chosen to maximize the probability of a successful outcome, i.e., a lattice very near a local optimum.

Lacking an analytic strategy to find the best annealing scheme for our purpose, we tested many options numerically. We focused on $n=8$, where, according to a 1998 conjecture \cite{agrell98}, there is a unique local optimum: the Gosset lattice $E_8$. We designed many thousands of $8$-dimensional lattices using the proposed SGD method and estimated how similar the obtained lattices were to $E_8$. However, comparing two lattices is not a trivial task, since any given lattice, including $E_8$, can be represented using infinitely many different generator matrices. Instead of comparing the generator matrices directly, we therefore numerically found the shortest nonzero lattice vectors in each of the obtained lattices. This was done using a modified version of the ``kissing number'' algorithm in \cite[Sec.~VI-B]{agrell02}. In the $E_8$ lattice, there are 240 shortest nonzero vectors, which have a squared norm of $2$ if the lattice is normalized to unit volume, and it is the unique $8$-dimensional lattice with this property \cite[pp.~22, 120--123]{conway99splag}. Our accuracy metric is therefore the mean square error of the squared Euclidean norms of the 240 shortest vectors, given that their true values should all be $2$. This metric is independent of basis changes and orthogonal transformation ($\bU$ and $\bR$ in \eqref{e:equiv}).

Three annealing schemes were tested, in which $\mu_t$ decreases linearly, quadratically, or exponentially with $t$. For each scheme, multiple step size parameters were tested. It turned out that the exponential decrease consistently showed better accuracy after a given number of iteration steps. We therefore apply an exponential annealing scheme in our proposed lattice construction algorithm, which is detailed in the next section.

\renewcommand{\algorithmicrequire}{\textbf{Input:}}
\renewcommand{\algorithmicensure}{\textbf{Output:}}
\begin{algorithm}[tb]
{\small
	\caption{Iterative lattice construction}
	\label{a:training}
	\begin{algorithmic}[1] %
		\REQUIRE Dimension $n$
		\ENSURE Generator matrix $\bB$
		\STATE $\bB \leftarrow \orth(\red(\gran(n,n)))$
		\STATE $V \leftarrow \prod_{i=1}^n [\bB]_{i,i}$
		\STATE $\bB \leftarrow V^{-1/n}\bB$
		\FOR{$t=0$ \TO $T-1$}
			\STATE $\mu \leftarrow \mu_0 \nu^{-t/(T-1)}$
			\STATE $\bz \leftarrow \uran(n)$
			\STATE $\by \leftarrow \bz-\clp(\bB,\bz\bB)$
			\STATE $\be \leftarrow \by\bB$
			\FOR{$i=1$ \TO $n$}
				\FOR{$j=1$ \TO $i-1$}
					\STATE $[\bB]_{i,j} \leftarrow [\bB]_{i,j}-\mu\, [\by]_i [\be]_j$
				\ENDFOR
				\STATE $[\bB]_{i,i} \leftarrow [\bB]_{i,i}-\mu
						\left( [\by]_i [\be]_i-\frac{\|\be\|^2}{n [\bB]_{i,i}} \right)$
			\ENDFOR
			\IF {($t \bmod \tr) = \tr-1$}
				\STATE $\bB \leftarrow \orth(\red(\bB))$
				\STATE $V \leftarrow \prod_{i=1}^n [\bB]_{i,i}$
				\STATE $\bB \leftarrow V^{-1/n}\bB$
			\ENDIF
		\ENDFOR
	\end{algorithmic}
}
\end{algorithm}

\begin{table*} \centering
\caption{Parameters of Algorithm \ref{a:training}.}
\label{t:parameters}
\begin{tabular}{ccccc}
\hline
Parameter & Description & Fast & Medium & Slow \\
\hline\hline
$\mu_0$ & Initial step size & 0.005 & 0.001 & 0.0005\\
$\nu$ & Ratio between initial and final step size & 200 & 500 & 1000 \\
$T$ & Number of steps & 1\,000\,000 & 10\,000\,000 & 100\,000\,000\\
$\tr$ & Reduction interval (should divide $T$) & 100 & 100 & 100 \\
\hline
\end{tabular}
\end{table*}

\section{Implementation} \label{s:implementation}

In this section, we present the lattice construction algorithm from a purely practical perspective. The theory developed in the previous section is necessary to fully understand its operation, but not to implement it.

The iterative algorithm is detailed in Algorithm \ref{a:training}. Its operation is controlled by four parameters, which are summarized in Tab.~\ref{t:parameters}. As usual, there is a trade-off between time and quality. Suggested sets of parameters are listed in the last three table columns, of which the last one yields the most accurate results at the cost of a higher computational complexity. Generally speaking, one should shift from ``Fast'' to ``Medium'' to ``Slow'' with increasing dimension to achieve the same accuracy, or for improved accuracy at a fixed dimension.
 
The utility functions $\gran$, $\uran$, $\clp$, $\red$, and $\orth$, on which the algorithm relies, are described in the following paragraphs.

The function $\uran(n)$ returns $n$ random real numbers, which are uniformly distributed in $[0,1)$ and independent of each other. Interpreted as a vector, $\uran(n)$ returns a random point uniformly distributed in an $n$-dimensional hypercube. We use the \emph{permuted congruential generator} \cite{oneill14}, which is well documented and fulfills advanced tests of randomness.\footnote{This is not always the case for the built-in random number generators in various programming languages.} Furthermore, $\gran(n,m)$ returns an $n\times m$ matrix of random independent real numbers, each with a Gaussian zero-mean, unit-variance distribution. We use the Gaussian random number generator by Paley and Wiener \cite[Eqs.~(37.01), (37.03)]{paley34}.\footnote{The method has later become known as the Box--Muller transform.}

The \emph{closest lattice point} function $\clp(\bB,\bx)$ finds the point in the lattice generated by $\bB$ that is closest to $\bx$. The output is not the lattice point itself but rather its integer coordinates $\argmin_{\bu\in\Z^n} \|\bx-\bu\bB\|^2$. We use \cite[Algorithm 5]{ghasemmehdi11}, which is the fastest general closest point search algorithm known to us. It applies to square, lower-triangular generator matrices with positive diagonal elements, which is exactly how lattices are represented in Algorithm \ref{a:training}.

The \emph{reduction} function $\red(\bB)$ returns another generator matrix for the lattice generated by $\bB$, in which the rows (basis vectors) are shorter and more orthogonal to each other than in $\bB$. If no improved generator matrix is found, $\bB$ is returned unchanged. A fast and popular algorithm for the purpose is the \emph{Lenstra--Lenstra--Lov\'asz algorithm} \cite[Fig.~1]{lenstra82}, which we apply in this work. In the context of \eqref{e:equiv}, reduction corresponds to finding a suitable $\bU$.

The \emph{orthogonal transformation} function $\orth(\bB)$ rotates and reflects an arbitrary generator matrix into a square, lower-triangular form with positive diagonal elements. This corresponds to finding a new coordinate system for the lattice, in which the first $i$ unit vectors span the subspace of the first $i$ basis vectors, for $i=1,\ldots,n$. We implement this function by Cholesky-decomposing \cite[Sec.~14.5]{harville97} the Gram matrix $\bA=\bB \bB^\T$. In the context of \eqref{e:equiv}, orthogonal transformation corresponds to right-multiplying the generator matrix by a matrix $\bR$ with orthonormal rows.

After Step 13, we apply a sanity check that $[\bB]_{i,i}$ is still positive (not shown in Algorithm \ref{a:training}). If any diagonal element is zero or negative, then the algorithm is aborted and the construction has failed, which suggests that the parameters $\mu_0$, $\tr$, or both may have been set too large. We encountered this problem during algorithm development, but never when the parameters were chosen as in Tab.~\ref{t:parameters}.

Steps 15--19 can be omitted under some conditions (e.g., small dimensions $n$ and small initial step sizes $\mu_0$), but we do not recommend it. The usefulness of reduction is discussed in Sec.~\ref{s:update}.

Steps 2--3 and 17--18 normalize the lattice to unit volume, which is sometimes necessary to avoid numerical underflow in \eqref{e:vol} for large-scale problems (large $n$ and $T$). For small- and medium-scale problems, normalizing $\bB$ does not affect the operation of the algorithm or the resulting NSM.

\section{Alternative algorithms} \label{s:algorithms}

Algorithm \ref{a:training} is not the only way to numerically optimize lattice quantizers. In this section, we review all alternative algorithms that we are aware of. We have implemented them all and confirmed that they work, but we have not compared their performance in sufficient detail to appoint a winner. A rigorous comparison would require careful meta-optimization to find the most efficient algorithm parameters, averaging over many trials in each dimension. This was done for Algorithm \ref{a:training} in Sec.~\ref{s:implementation} but not for the alternative algorithms reviewed in this section.

\subsection{The Agrell--Eriksson algorithm}

The first, and so far only, published algorithm for lattice quantizer optimization was proposed by Agrell and Eriksson in 1998 \cite{agrell98}. It serves as a benchmark for the present study. Like Algorithm \ref{a:training}, the 1998 benchmark restricts the generator matrix to a square, lower-triangular form and updates the elements by SGD.  The first $n-1$ dimensions are updated in the direction of the negative gradient of the unnormalized second moment, whereas the last dimension (controlled by a single element in the bottom-right corner) in every iteration is set to maintain a unit volume. An example serves to highlight how this strategy differs from using SGD with the NSM gradient.

\begin{example} \label{ex:4d}
We consider the first iteration when optimizing a four-dimensional lattice. For example, let $n=4$, $t=0$, and $\bB = \bB_0 = \bI_4$. Assume that the first random vector is $\bz = \bz_0 = [0.6, 0.6, 0.0, 0.0]$. The closest point problem \eqref{e:uhat} is solved by $\uhat = [1,1,0,0]$. We subsequently obtain $\be = (\bz-\uhat)\bB = [-0.4, -0.4, 0.0, 0.0]$, $\|\be\|^2 = 0.32$, $V=1$, and $g(\bB,\bz) = 0.08$.

If the benchmark method \cite[Tab.~I]{agrell98} is applied, then the updated generator matrix is
\eqlab{e:ex2}{
\bB_1' &=
\begin{bmatrix}
1-0.32\epsilon & 0 & 0 & 0 \\
-0.32\epsilon & 1-0.32\epsilon & 0 & 0 \\
0 & 0 & 1 & 0 \\
0 & 0 & 0 & \frac{1}{(1-0.32\epsilon)^2}
\end{bmatrix} \notag\\
&= \bB_0 - \epsilon
\begin{bmatrix}
0.32 & 0 & 0 & 0 \\
0.32 & 0.32 & 0 & 0 \\
0 & 0 & 0 & 0 \\
0 & 0 & 0 & -0.64
\end{bmatrix}
+ O(\epsilon^2) \notag\\
&= \bB_0 - \bareps
\begin{bmatrix}
0.3780 & 0 & 0 & 0 \\
0.3780 & 0.3780 & 0 & 0 \\
0 & 0 & 0 & 0 \\
0 & 0 & 0 & -0.7559
\end{bmatrix}
+ O(\bareps^2),
}
where we in the last step substituted $\epsilon = 1.181\bareps$, so that $\bareps$ multiplies a matrix with unit Frobenius norm.
With the $\bB_1'$ and the same $\bz$ as before, we obtain $\be = [-0.4+0.256\bareps, -0.4+0.256\bareps, 0, 0]$, $V=1$, and $g(\bB,\bz) =
0.08-0.0907\bareps+0.0286\bareps^2$.
The negative coefficient of $\bareps$ shows that $g(\bB,\bz)$ decreases for small values of $\bareps$, as expected. 

If instead Algorithm \ref{a:training} is applied in the same scenario, the updated generator matrix is from \eqref{e:dgz2}--\eqref{e:update}
\eqlab{e:ex1}{
\bB_1 &= \bB_0 - \epsilon
\begin{bmatrix}
0.04 & 0 & 0 & 0 \\
0.08 & 0.04 & 0 & 0 \\
0 & 0 & -0.04 & 0 \\
0 & 0 & 0 & -0.04
\end{bmatrix} \notag\\
&= \bB_0 - \bareps
\begin{bmatrix}
0.3536 & 0 & 0 & 0 \\
0.7071 & 0.3536 & 0 & 0 \\
0 & 0 & -0.3536 & 0 \\
0 & 0 & 0 & -0.3536
\end{bmatrix},
}
where we substituted $\epsilon = 8.839\bareps$ so that $\bareps$ in \eqref{e:ex1} as well as in \eqref{e:ex2} multiplies a matrix with unit Frobenius norm.
After the update, we have $\be = (\bz-\uhat)\bB_1 = [-0.4+0.4243\bareps, -0.4+0.1414\bareps, 0, 0]$, $V = (1-0.125\bareps^2)^2$, and from \eqref{e:gz}
\eqlab{e:gseries}{
g(\bB,\bz) &= \frac{0.32-0.4525\bareps+0.2\bareps^2}{4(1-0.125\bareps^2)} \notag\\
&= 0.08-0.1131\bareps+0.06\bareps^2+O(\bareps^3)
}
for the given $\bz = \bz_0$. Again, the coefficient of $\bareps$ is negative, indicating a decreasing $g(\bB,\bz)$.
\hfill$\bigtriangleup$\end{example}

The convergence rate for small step sizes $\bareps$ is governed by the linear term, which in this example is $-0.0907\bareps$ and $-0.1131\bareps$ for the two update rules, respectively. Evidently, $g(\bB,\bz)$ decreases faster with Algorithm \ref{a:training} than with the benchmark, which illustrates that the latter does not follow the gradient of the NSM. 
Therefore, the benchmark generally needs more iterations for the same accuracy than Algorithm \ref{a:training}, which we have also observed numerically. For other values of $\bB$ and $\bz$ than the ones considered in Example \ref{ex:4d}, the difference between the $g(\bB,\bz)$ values obtained with the two update rules can be larger or smaller, but Algorithm \ref{a:training} always reduces $g(\bB,\bz)$ more than the benchmark does.
Indeed, the coefficient of $\bareps$ in \eqref{e:gseries} is the smallest (most negative) possible and represents the direction of steepest descent subject to the lower-triangular constraint, since the elements of the right-hand matrix in \eqref{e:ex1} are the partial derivatives in \eqref{e:dgz2}.

Another indication that the benchmark performs suboptimally is that the third and fourth dimensions are updated differently in \eqref{e:ex2}, even though the input data in this (artificial) example is fully symmetric with respect to these two dimensions. This is because the  fourth dimension is used to maintain a constant volume. With Algorithm \ref{a:training}, in contrast, the third and fourth dimensions are updated equally, as evident from \eqref{e:ex1}.

In the benchmark, as presented in \cite{agrell98}, the initial generator matrix was set to the identity matrix. The motivation was that $G$ as a function of $\bB$ has a saddle point at $\bI_n$, from which subsequent iterations can bring $\bB_t$ down into one of several different ``valleys.'' Thereby, the algorithm is able to find different local minima in subsequent runs. However, although it seems intuitively likely and previous results show that it is the case in low dimensions \cite{agrell98}, we cannot guarantee that there is a downhill route from the identity matrix into a global minimum. What if the global minimum lies on the other side of another saddle point, which the SGD algorithm does not have the momentum to overcome? To avoid this potential risk, the initial generator matrix is chosen randomly in Algorithm \ref{a:training}.

Other minor differences between the benchmark and Algorithm \ref{a:training} are in the choices of step size annealing scheme (linear in the benchmark and exponential in Algorithm \ref{a:training}) and closest point search algorithm (\!\!\cite{agrell02} in the benchmark and \cite{ghasemmehdi11} in Algorithm \ref{a:training}).

\subsection{Full-matrix stochastic gradient descent} \label{s:fullmatrix}

In Algorithm \ref{a:training}, the generator matrix is constrained to a lower-triangular structure, and the updates follow the negative gradient of the NSM subject to this constraint  for complexity reasons. A different, and perhaps more natural, approach is to update an unconstrained $n\times n$ generator matrix according to the negative gradient of the NSM with respect to all matrix elements. Even if the initial matrix would be lower-triangular, the NSM gradient would soon bring it away from that structure. An optimization algorithm based on this principle can be defined as follows.

Let
\eqlab{e:gzfull}{
g(\bB,\bz) = \frac{1}{n}|\det\bB|^{-2/n} \|\be\|^2,
}
where, in contrast to the analogous equation \eqref{e:gz}, the absolute value is needed because $\det\bB$ may here be positive or negative. As in Algorithm \ref{a:training}, $\bz$ is uniform in $[0,1)^n$, $\by = \bz-\uhat$ with $\uhat$ given by \eqref{e:uhat}, and $\be=\by\bB$. Again we neglect the zero-measure set of $\bz$ for which \eqref{e:uhat} does not have a unique solution.

The gradient of \eqref{e:gzfull} with respect to $\bB$ is
\eqlab{e:dgzfull}{
\nabla g(\bB,\bz) =
  \frac{1}{n}|\det\bB|^{-2/n} \nabla\|\be\|^2 + \frac{1}{n}\|\be\|^2\nabla|\det\bB|^{-2/n}
.}
The two gradients on the right-hand side of \eqref{e:dgzfull} are
\eq{
\nabla\|\be\|^2 &= \nabla (\by \bB \bB^\T \by^\T) \notag\\
  &= 2\by^\T \by\bB \notag\\
  &= 2\by^\T\be, \label{e:de2full} \\
\nabla|\det\bB|^{-2/n} &= -\frac{2}{n}|\det\bB|^{-2/n-1} \nabla|\det\bB| \notag\\
  &= -\frac{2}{n}|\det\bB|^{-2/n-1} (\sgn\det\bB) \nabla\det\bB \notag\\
  &= -\frac{2}{n}|\det\bB|^{-2/n-1} (\sgn\det\bB) (\adj\bB)^\T \label{e:dv}
,}
where $\sgn$ denotes the signum function and $\adj$ the adjugate or adjoint matrix%
\footnote{Not to be confused with the conjugate transpose, which is also sometimes called adjoint.}
\cite[Sec.~15.8]{harville97}.
Combining \eqref{e:dgzfull}, \eqref{e:de2full}, and \eqref{e:dv} yields after simplification
\eq{
\nabla g(\bB,\bz) &=
  \frac{2}{n} |\det\bB|^{-2/n} \left( \by^\T\be - \frac{\|\be\|^2}{n \det\bB} (\adj\bB)^\T \right) \notag \\
  &= \frac{2}{n} |\det\bB|^{-2/n} \left( \by^\T\be - \frac{\|\be\|^2}{n} (\bB^{-1})^\T \right)  
.}

As an SGD update rule, analogous to \eqref{e:muupdate}, we can use
\eqlab{e:updatefull}{
\bB_{t+1} &= \bB_t - \mu_t \frac{n |\det\bB_t|^{2/n}}{2} \nabla g(\bB_t,\bz_t) \notag \\
&= \bB_t - \mu_t \left( \by_t^\T\be_t - \frac{\|\be_t\|^2}{n} (\bB_t^{-1})^\T \right)
.}
The result depends on the choice of initial lattice, number of iterations, annealing scheme for $\mu_t$, and frequency of reduction and renormalization. With our preliminary implementation, this algorithm appears to need fewer iterations than Algorithm \ref{a:training} to converge to a locally optimal lattice quantizer with the same accuracy, but every iteration is more complex. This is because general-purpose closest lattice point search algorithms need a triangular generator matrix \cite{agrell02, ghasemmehdi11}. Thus, $\bB$ needs to be converted to lower-triangular form by Cholesky or QR decomposition in every iteration. The inverse in \eqref{e:updatefull} also adds complexity compared with \eqref{e:dgz2}--\eqref{e:muupdate}.

The more complex iterations can partly be mitigated by treating $p$ independent samples $\bz_{t,i} \in [0,1)^n$ for $i=1,\ldots,p$ in a batch:
\eq{
\by_{t,i} &= \bz_{t,i} - \argmin_{\bu\in\Z^n} \|(\bz_{t,i}-\bu)\bB_t\|^2, \\
\be_{t,i} &= \by_{t,i}\bB_t, \\
\bB_{t+1} &= \bB_t - \frac{\mu_t}{p}\sum_{i=1}^p\left( \by_{t,i}^\T\be_{t,i} - \frac{\|\be_{t,i}\|^2}{n} (\bB_t^{-1})^\T \right)
.}
With this modification, the triangular decomposition and matrix inverse need to be calculated only for every $p$th closest point search.

We do not know if this SGD variant performs better than Algorithm \ref{a:training} if both are configured for the same execution time. It depends on the algorithm parameters, the employed hardware and software, and the dimension.

\subsection{Whitening the quantization noise} \label{s:whitening}

In 1996, Zamir and Feder proved that the optimal lattice quantizer in every dimension has white quantization noise; i.e., a covariance matrix 
\eqlab{e:covariance}{
  \bC = \frac{1}{V} \int_\Omega \bx^\T \bx \dx
}
proportional to the identity \cite{zamir96}. Furthermore, if a lattice $\cL$ has a covariance matrix $\bC$ that is not proportional to the identity, then $\cL \bC^{-1/2}$ has a lower NSM than $\cL$. The whiteness property was generalized to locally optimal lattices in \cite{agrell23} and applied to optimize a parametric lattice in \cite{pook-kolb24arxiv}.
The covariance matrix can be empirically estimated as $\hat{\bC} = \bE^\T \bE/p$, where $\bE$ is a $p\times n$ vector whose rows are independent random vectors, uniformly distributed in $\Omega$, and $p\ge n$. The generation of uniform random vectors in $\Omega$ is discussed in Sec.~\ref{s:estimation}.

A simple lattice optimization algorithm follows straightforwardly from the white noise property. The idea is to repeatedly estimate the covariance matrix and update the generator matrix. Every update consists of computing
\eq{
\bB_{t+1} = \bB_t \left( \frac{\bE_t^\T \bE_t}{p} \right)^{-1/2},
}
where $\bE_t$ is composed of $p$ vectors in the Voronoi region of the lattice generated by $\bB_t$, possibly followed by reduction and renormalization. Much fewer iterations are needed than with any of the previously described algorithms, but every iteration is more complex, involving $p$ closest point searches. Like in the previous algorithm, the generator matrix needs to be rotated into lower-triangular form in every iteration to enable closest point search.

Good results were obtained when $p$ increases with $t$. Apart from this simple observation, we leave the meta-optimization of the algorithm parameters as an open question.

\subsection{Simulated annealing}

\emph{Simulated annealing} is a classical heuristic in numerical optimization. The idea is to accept suboptimal updates with some probability, so that the algorithm will sometimes move ``uphill,'' away from local minima.%
\footnote{The annealing scheme used to adjust the step size in Sec.~\ref{s:stepsize} is not simulated annealing in this sense.}
The probability is decreased with time as the optimization progresses and is sometimes referred to as ``temperature.'' Simulated annealing has been successfully applied in the design of several types of codes and quantizers \cite{elgamal87, farvardin90}, but not lattices, as far as we know.

All algorithms mentioned in this paper can be modified to include some aspects of simulated annealing. We propose to introduce it in the closest lattice point search, which consumes a large share of the computational resources in all considered optimization algorithms. Such search methods \cite{agrell02, ghasemmehdi11} work by enumerating lattice points inside a sphere and reducing the size of the sphere every time a lattice point is found, until the sphere is empty. Simulated annealing can be implemented in this context by aborting the search with some probability every time a new lattice point is found. Thereby we achieve not only the desired erratic movement but also reduced computational complexity. If the probability, or temperature, is $100\,\%$, then the returned vector is the so-called \emph{Babai point,} which is found very fast \cite{babai86,agrell02}, and if it is $0\,\%$, then the search method reverts to its normal, optimal behavior.

\section{Identification of Exact Lattices} \label{s:identification}

When a numerical lattice optimization algorithm terminates, the result is a generator matrix with numerical entries. Such matrices can be used to numerically estimate various lattice parameters, such as the NSM, packing radius, and covering radius. They do not immediately offer any geometrical insights about for example symmetries or algebraic construction methods,
nor do they enable any efficient search strategies such as the closest point algorithms in \cite[Ch.~20]{conway99splag}, but they can pave the way for such progress if their underlying exact forms can be identified. Therefore, the algorithm's outputs should if possible be refined. We do this in three steps: first, computing and visualizing the theta series; second, replacing the numerically defined lattice with a similar but exact lattice, whose theta series has a certain desired form; and third, checking the obtained exact lattice for consistency.

\begin{figure} \centering
\begin{tikzpicture}
\sffamily\small
\begin{axis}[
width=\columnwidth, height=6cm,
xmin=0, xmax=5.3,
minor x tick num=1,
xlabel={$r^2$},
xlabel near ticks,
ymode=log,
ymin=0.4, ymax=15000,
ylabel={$N(\bB,r)$},
ylabel near ticks,
log ticks with fixed point,
ytick distance=10,
yticklabel style={/pgf/number format/1000 sep=\,}, %
legend style={at={(.96,.06)}, anchor=south east, legend cell align=left, font=\footnotesize}
]
\legend{$t=0$, $t=100\,000$, $t=300\,000$, $t=1\,000\,000$, $t=10\,000\,000$}
\addplot[blue!32!white, line width=0.8, densely dashed] table{logtheta1.txt};
\addplot[blue!49!white, line width=0.8, densely dashed] table{logtheta2.txt};
\addplot[blue!66!white, line width=0.8, densely dashed] table{logtheta3.txt};
\addplot[blue!83!white, line width=0.8, densely dashed] table{logtheta4.txt};
\addplot[blue, line width=0.8] table{logtheta5.txt};
\end{axis}
\end{tikzpicture}
\caption{The evolution of a $10$-dimensional lattice during numerical optimization.}
\label{f:theta10}
\end{figure}
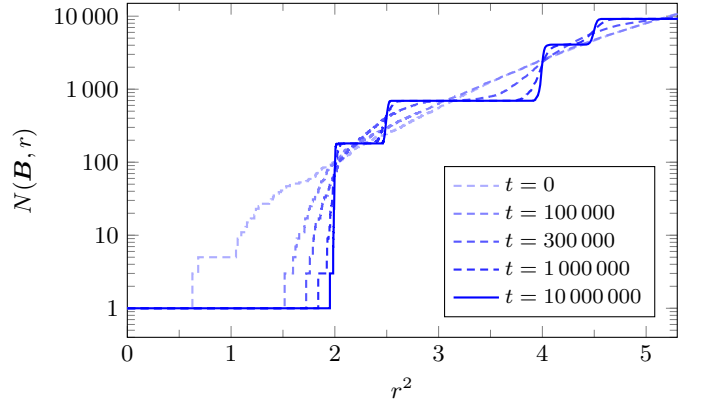

\subsection{Theta Image: Visualizing Lattice Structure} \label{s:theta}

To monitor the progression during optimization and visualize the convergence, we find it informative to study the cumulative distribution of lattice point norms
\eq{
N(\bB,r) = \left| \{\bu \in \Z^n \colon \| \bu\bB \| \le r \} \right|
.}
The function can be numerically computed for any $\bB$ and moderate values of $r$ by trivial modifications of the kissing number algorithm in \cite[Sec.~VI-B]{agrell02}. We call a plot of $N(\bB,r)$ versus $r^2$ a \emph{theta image,} because it illustrates the \emph{theta series} \eqref{e:theta} of the lattice generated by $\bB$. For any lattice, $N(\bB,r) = 1$ for $0<r<2\rho$ and $N(\bB,2\rho) = 1 + \tau$, where, as defined in Sec.~\ref{s:lattices}, $\rho$ is the packing radius and $\tau$ is the kissing number.

Fig.~\ref{f:theta10} illustrates how the theta image of a lattice evolves during the iterations of Algorithm \ref{a:training} and converges to a local minimum. The dimension was set to $n=10$ and the optimization parameters were taken from the column ``Medium'' in Tab.~\ref{t:parameters}. As the algorithm progresses, the theta image approaches a staircase-like curve, composed of very steep and very flat sections, which indicates that the lattice points move towards a small number of spherical shells around the origin. We have observed this behavior in all dimensions, provided that the optimization parameters are well chosen. The height of each vertical step indicates the number of lattice points in the corresponding shell and the horizontal location of the step indicates its squared radius.

The convergence towards discrete shells around the origin is not surprising, given that good lattices usually have a high degree of symmetry. Known theta series, which are listed for many classical lattices in \cite[Ch.~4]{conway99splag}, confirm that many lattice points have equal norms in such lattices.

\begin{figure} \centering
\begin{tikzpicture}
\sffamily\small
\begin{axis}[
width=\columnwidth, height=6cm,
xmin=0, xmax=5.3,
minor x tick num=1,
xlabel={$r^2$},
xlabel near ticks,
ymode=log,
ymin=0.4, ymax=15000,
ylabel={$N(\bB,r)$},
ylabel near ticks,
ytick distance=10,
log ticks with fixed point,
yticklabel style={/pgf/number format/1000 sep=\,}, %
]
\addplot[orange, line width=1.2, densely dashdotted] table[row sep=\\]{
	0	1\\
	2	1\\
	2	181\\
	2.5	181\\
	2.5	693\\
	4	693\\
	4	4073\\
	4.5	4073\\
	4.5	9193\\
	5.4	9193\\
};
\end{axis}
\end{tikzpicture}
\caption{The exact theta image of $D^+_{10}$, towards which the curves in Fig.~\ref{f:theta10} converge.}
\label{f:theta10Dplus}
\end{figure}
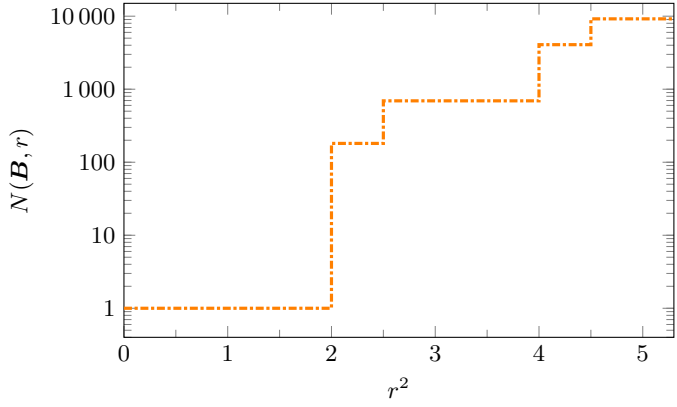

\subsection{From Approximate to Exact Theta Image} \label{s:exact}

The observation that numerically optimized lattices tend to have many lattice vectors with similar norms suggests a systematic approach to the generation of exact lattice representations. In this section, we describe a technique to perturb the Gram matrix so that all lattice points whose norms are numerically \emph{almost} equal become \emph{exactly} equal.

The squared norm of a lattice vector $\bu\bB$ is the \emph{quadratic form} $\|\bu\bB\|^2 = \bu\bA\bu^\T$, where $\bu \in \Z^n$. If $\bA$ is not known exactly but only through an approximation $\tA$, then we can use the theta image of $\tA$ to identify a set of integer vectors $\bu_1,\ldots, \bu_M$ for which $\bu_1\tA\bu_1^\T \approx \cdots \approx \bu_M\tA\bu_M^\T$. Thereafter we replace $\tA$ with an unknown, generic $\bA$ of the same dimension and solve the equations $\bu_1\bA\bu_1^\T = \cdots = \bu_M\bA\bu_M^\T$ with the same integer vectors. Additional equations can be obtained from other vertical steps in the theta image. We do not impose a certain value for the norms, only that they should be equal to each other. The first diagonal element of $\bA$ is set to $1$ to fix the scale and $\bA$ is symmetric, giving a system of equations with $n(n+1)/2-1$ unknowns.

The system of equations is linear and can be solved exactly using standard methods. One of three possible outcomes can occur: one solution, multiple solutions, or no solution. If a single solution occurs, then it is presumably a locally optimal lattice quantizer and the investigation can proceed to the next stage, which is to characterize its properties and compare it with known lattices. If multiple solutions occur, then the system of equations was underdetermined and it may help to extract equations from additional vertical steps in the theta image. If these additional equations do not reduce the number of solutions, it may be because the symmetry group of the underlying exact lattice is separable into disjoint groups involving separate lower-dimensional rotations. The solution may still identify lower-dimensional sublattices and thus reduce the number of unknowns from $n(n+1)/2-1$, even though at least one unknown remains. Such \emph{parametric lattices} require additional numerical and analytical methods to be identified exactly \cite{pook-kolb24arxiv}. The smallest dimension for which this occurs is $n=9$, where a single parameter cannot be identified by symmetry considerations \cite{agrell98,allen21adp}. The third and final outcome is that no solution is returned. Then the system was overdetermined and a plausible explanation is that what looks like a single vertical step in the theta image may theoretically consist of two steps at different but similar norms. We have not yet encountered this outcome in any practical trials.

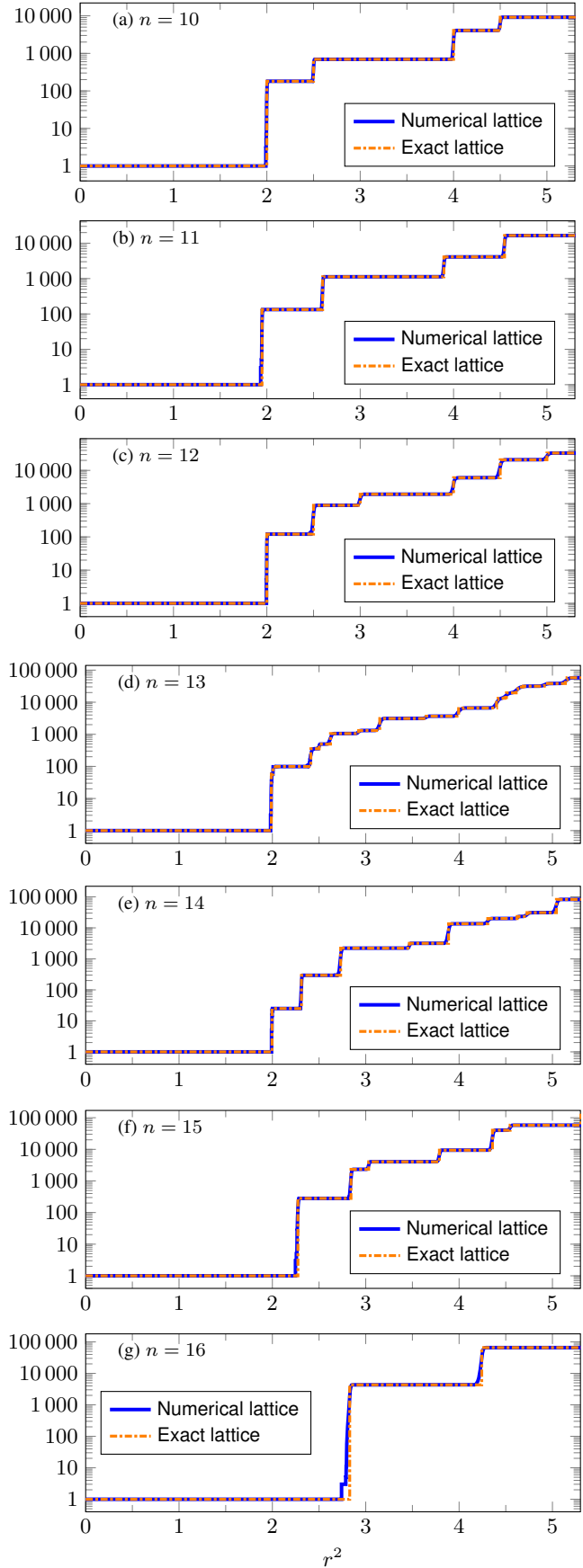
\begin{figure} \centering
\begin{tikzpicture}
\sffamily\small
\begin{axis}[ %
width=\columnwidth, height=4.2cm,
xmin=0, xmax=5.3,
minor x tick num=1,
ymode=log,
ymin=0.4, ymax=22000,
log ticks with fixed point,
ytick distance=10,
yticklabel style={/pgf/number format/1000 sep=\,}, %
legend style={at={(.96,.08)}, anchor=south east, legend cell align=left, font=\footnotesize}
]
\legend{Numerical lattice, Exact lattice}
\addplot[blue, line width=1.5] table{logth10.txt};
\addplot[orange, line width=1.2, densely dashdotted] table[row sep=\\]{
	0	1\\
	2	1\\
	2	181\\
	2.5	181\\
	2.5	693\\
	4	693\\
	4	4073\\
	4.5	4073\\
	4.5	9193\\
	5.4	9193\\
};
\node[anchor=west] at (rel axis cs:.05,.9) {\footnotesize $\textrm{(a)} \; n=10$};
\end{axis}
\end{tikzpicture}
\\[1ex]
\begin{tikzpicture}
\sffamily\small
\begin{axis}[ %
width=\columnwidth, height=4.2cm,
xmin=0, xmax=5.3,
minor x tick num=1,
ymode=log,
ymin=0.4, ymax=43000,
log ticks with fixed point,
ytick distance=10,
yticklabel style={/pgf/number format/1000 sep=\,}, %
legend style={at={(.96,.08)}, anchor=south east, legend cell align=left, font=\footnotesize}
]
\legend{Numerical lattice, Exact lattice}
\addplot[blue, line width=1.5] table{logth11.txt};
\addplot[orange, line width=1.2, densely dashdotted] table[row sep=\\]{
	0	1\\
	1.94837	1\\
	1.94837	133\\
	2.59783	133\\
	2.59783	1123\\
	3.89674	1123\\
	3.89674	4093\\
	4.5462	4093\\
	4.5462	16501\\
	5.4	16501\\
};
\node[anchor=west] at (rel axis cs:.05,.9) {\footnotesize $\textrm{(b)} \; n=11$};
\end{axis}
\end{tikzpicture}
\\[1ex]
\begin{tikzpicture}
\sffamily\small
\begin{axis}[ %
width=\columnwidth, height=4.2cm,
xmin=0, xmax=5.3,
minor x tick num=1,
ymode=log,
ymin=0.4, ymax=90000,
log ticks with fixed point,
ytick distance=10,
yticklabel style={/pgf/number format/1000 sep=\,}, %
legend style={at={(.96,.08)}, anchor=south east, legend cell align=left, font=\footnotesize}
]
\legend{Numerical lattice, Exact lattice}
\addplot[blue, line width=1.5] table{logth12.txt};
\addplot[orange, line width=1.2, densely dashdotted] table[row sep=\\]{
	0	1\\
	2	1\\
	2	121\\
	2.5	121\\
	2.5	889\\
	3	889\\
	3	1913\\
	4	1913\\
	4	6017\\
	4.5	6017\\
	4.5	20865\\
	5	20865\\
	5	33153\\
	5.4	33153\\
};
\node[anchor=west] at (rel axis cs:.05,.9) {\footnotesize $\textrm{(c)} \; n=12$};
\end{axis}
\end{tikzpicture}
\\[1ex]
\begin{tikzpicture}
\sffamily\small
\begin{axis}[ %
width=\columnwidth, height=4.2cm,
xmin=0, xmax=5.3,
minor x tick num=1,
ymode=log,
ymin=0.4, ymax=140000,
log ticks with fixed point,
ytick distance=10,
yticklabel style={/pgf/number format/1000 sep=\,}, %
legend style={at={(.96,.08)}, anchor=south east, legend cell align=left, font=\footnotesize}
]
\legend{Numerical lattice, Exact lattice}
\addplot[blue, line width=1.5] table{logth13.txt};
\addplot[orange, line width=1.2, densely dashdotted] table{
0	1
1.9888	1
1.9888	57
2.00608	57
2.00608	97
2.04884	97
2.04884	99
2.41202	99
2.41202	355
2.50101	355
2.50101	495
2.62269	495
2.62269	1055
2.92423	1055
2.92423	1311
3.15031	1311
3.15031	3103
3.64712	3103
3.64712	3663
3.9776	3663
3.9776	4083
3.99488	4083
3.99488	6323
4.01217	6323
4.01217	6413
4.03764	6413
4.03764	6525
4.05493	6525
4.05493	6605
4.40082	6605
4.40082	11981
4.4181	11981
4.4181	13261
4.48981	13261
4.48981	13933
4.50709	13933
4.50709	19533
4.61149	19533
4.61149	25293
4.62878	25293
4.62878	29773
4.68695	29773
4.68695	31565
4.91303	31565
4.91303	36941
4.93032	36941
4.93032	38221
5.13911	38221
5.13911	48973
5.1564	48973
5.1564	57933
5.4	57933
};
\node[anchor=west] at (rel axis cs:.05,.9) {\footnotesize $\textrm{(d)} \; n=13$};
\end{axis}
\end{tikzpicture}
\\[1ex]
\begin{tikzpicture}
\sffamily\small
\begin{axis}[ %
width=\columnwidth, height=4.2cm,
xmin=0, xmax=5.3,
minor x tick num=1,
ymode=log,
ymin=0.4, ymax=220000,
log ticks with fixed point,
ytick distance=10,
yticklabel style={/pgf/number format/1000 sep=\,}, %
legend style={at={(.96,.08)}, anchor=south east, legend cell align=left, font=\footnotesize}
]
\legend{Numerical lattice, Exact lattice}
\addplot[blue, line width=1.5] table{logth14.txt};
\addplot[orange, line width=1.2, densely dashdotted] table{
0	1
1.99599	1
1.99599	25
2.31126	25
2.31126	295
2.73144	295
2.73144	2215
3.46689	2215
3.46689	3175
3.88707	3175
3.88707	13543
3.99197	13543
3.99197	13567
4.30724	13567
4.30724	20047
4.62251	20047
4.62251	23287
4.72742	23287
4.72742	30967
5.04269	30967
5.04269	82807
5.4	82807

};
\node[anchor=west] at (rel axis cs:.05,.9) {\footnotesize $\textrm{(e)} \; n=14$};
\end{axis}
\end{tikzpicture}
\\[1ex]
\begin{tikzpicture}
\sffamily\small
\begin{axis}[ %
width=\columnwidth, height=4.2cm,
xmin=0, xmax=5.3,
minor x tick num=1,
ymode=log,
ymin=0.4, ymax=170000,
log ticks with fixed point,
ytick distance=10,
yticklabel style={/pgf/number format/1000 sep=\,}, %
legend style={at={(.96,.08)}, anchor=south east, legend cell align=left, font=\footnotesize}
]
\legend{Numerical lattice, Exact lattice}
\addplot[blue, line width=1.5] table{logth15.txt};
\addplot[orange, line width=1.2, densely dashdotted] table{
0	1
2.27357	1
2.27357	281
2.84197	281
2.84197	2329
3.03143	2329
3.03143	4039
3.78929	4039
3.78929	9415
4.35769	9415
4.35769	40135
4.54715	40135
4.54715	58055
5.30501	58055
5.30501	120455
5.4	120455
};
\node[anchor=west] at (rel axis cs:.05,.9) {\footnotesize $\textrm{(f)} \; n=15$};
\end{axis}
\end{tikzpicture}
\\[1ex]
\begin{tikzpicture}
\sffamily\small
\begin{axis}[ %
width=\columnwidth, height=4.2cm,
xmin=0, xmax=5.3,
minor x tick num=1,
xlabel={$r^2$},
xlabel near ticks,
ymode=log,
ymin=0.4, ymax=180000,
log ticks with fixed point,
ytick distance=10,
yticklabel style={/pgf/number format/1000 sep=\,}, %
legend style={at={(.03,.50)}, anchor=west, legend cell align=left, font=\footnotesize}
]
\legend{Numerical lattice, Exact lattice}
\addplot[blue, line width=1.5] table{logth16.txt};
\addplot[orange, line width=1.2, densely dashdotted] table[row sep=\\]{
	0	1\\
	2.82843	1\\
	2.82843	4321\\
	4.24264	4321\\
	4.24264	65761\\
	5.4	65761\\
};
\node[anchor=west] at (rel axis cs:.05,.9) {\footnotesize $\textrm{(g)} \; n=16$};
\end{axis}
\end{tikzpicture}
\caption{Theta images of numerically optimized lattices in dimensions $10$--$16$ and of the exact lattices towards which these optimized lattices converge.}
\label{f:theta10-16}
\end{figure}

\subsection{Validation and Identification}

An important consistency check is now to compute the theta series of the obtained exact lattice, including more terms than those considered in the equation-solving framework described in Sec.~\ref{s:exact}. If those additional terms match vertical steps in the theta image of the numerically optimized lattice, then it is a strong indication that its exact form has been correctly determined. Furthermore, the theta series can be compared with other theta series found numerically, analytically, or in the literature, to help identify equivalences.

We return
to the $10$-dimensional lattice example of Sec.~\ref{s:theta}. When the optimization algorithm terminates, it has $180$ points with squared norms between $1.952$ and $2.043$ and $512$ points with squared norms between $2.460$ and $2.544$. The lattice has no points with squared norms between $2.043$ and $2.460$, and no nonzero points with squared norm less than $1.952$. The two point sets, which are visualized in the first two vertical steps in Fig.~\ref{f:theta10}, provide together $179+511 = 690$ linear equations in $54$ unknowns. The system of equations has a unique solution, apart from an overall scale factor, which gives a Gram matrix with rational elements.
The mean square error between the elements of the numerical and exact Gram matrices is $7.0\cdot 10^{-4}$, when both matrices are scaled to determinant $1$.
When instead the ``Slow'' parameters in Tab.~\ref{t:parameters} were used, a mean square error of $7.4\cdot 10^{-5}$ was obtained.
These error values should be interpreted as mere indications, since they were not averaged over many trials.

The theta image of the resulting lattice, scaled to $\det\bA=1$, is shown in Fig.~\ref{f:theta10Dplus}. Indeed all vertical steps, not only the first two, agree perfectly with steps in Fig.~\ref{f:theta10}.
The theta series corresponding to Fig.~\ref{f:theta10Dplus} is
\eq{
  1+180q^2+512q^{5/2}+3380q^4+5120q^{9/2}+\cdots
.}
This result suggests that the obtained lattice might be $D^+_{10}$, which has exactly the same theta series \cite[p.~120]{conway99splag} and has been previously suggested as the best $10$-dimensional lattice quantizer \cite{agrell98}.

\section{Lattice Quantizers in Dimensions $10$--$16$} \label{s:poc}

As a proof of concept, lattices were numerically optimized in dimensions $10$--$16$. Algorithm \ref{a:training} was implemented on a high-performance computer cluster, running on $100$ cores in parallel.
Since the NSM is a nonconvex function of the elements of $\bB$ with multiple minima, we designed $100$ lattices in each dimension. The optimization parameters are listed under ``Slow'' in Tab.~\ref{t:parameters}.

The NSM $G$ of each lattice was numerically estimated as described in Sec.~\ref{s:estimation}. To compute $\hat G$ in \eqref{e:ghat}, lines $6$--$8$ of Algorithm \ref{a:training} were repeated many times (in this case, $10^8$ times), the obtained squared vector norms $\|\be\|^2$ were averaged, and the result was normalized. The variance $\hat\sigma^2$ of the NSM estimate $\hat G$ was estimated according to \eqref{e:sighat}. In each dimension, several lattices with the smallest $\hat G$ were selected for further analysis and possible identification.

For the selected lattices in each dimension $n=10,\ldots,16$, the theta image was computed and plotted as described in Sec.~\ref{s:theta}. It turned out that these lattices have almost identical theta images, which suggests that they represent equivalent lattices, although their Gram (and generator) matrices are quite different. Specifically, the number of hits into the best equivalence class ranged from $16$ (for $n=15$) to $96$ (for $n=16$). It therefore seems plausible that $100$ trials in each dimension is more than enough to find the optimal lattices, although there is no guarantee.

For the best lattices in each dimension, vertical steps in their theta images were identified visually and exploited to define a system of equations as described in Sec.~\ref{s:exact}. In five of the seven cases, the system of equations had a unique solution, which indicated the exact lattice towards which the optimization process had converged. In two cases, more advanced identification methods were needed \cite{pook-kolb24arxiv}.
The theta images of the obtained exact lattices matches those of the numerically optimized lattices excellently, as shown in Fig.~\ref{f:theta10-16}.
As sanity checks, we confirmed that the volume of the Voronoi region in each case is $V=\det \bB$ and that the normalized second moment matrix \cite{zamir96} is
\eq{
    \frac{1}{V} \int_\Omega \bx^T \bx \dx
    = GV^{2/n} \bI_{n}
.}
The former condition holds for any lattice (see Sec.~\ref{s:lattices}) and the latter holds for any locally optimal $n$-dimensional lattice quantizer \cite{agrell23}.
We conjecture that all seven lattices are optimal in their respective dimensions.

The next step was to investigate if the found lattices are equivalent to any known lattices. The theta series proved useful for this purpose too, because having identical theta series is a necessary (but not sufficient) condition for equivalence between two lattices \cite[pp.~xxix, 47]{conway99splag}. We computed the theta series of approximately $110$ known lattices in dimensions $10$--$16$, all normalized to unit volume, and compared with the theta series of the found lattices. If an identical theta series was found, we verified the equivalence by finding explicit matrices $\bU$ and $\bR$ that transformed one of the generator matrices into the other according to \eqref{e:equiv}. The results are summarized in Tab.~\ref{t:results}.

In dimensions $10$--$12$, the algorithm converged to the previously best known lattice quantizers, suggesting that these might indeed be optimal in their respective dimensions. For $n=10$, $D^+_{10}$ was proposed in \cite{agrell98} and its exact NSM was computed in \cite{dutour09}. For $n=11$, the Coxeter lattice $A_{11}^3$ was proposed and its exact NSM was computed in \cite{dutour09}. For $n=12$, a lattice constructed by gluing four copies of $D_6\times D_6$ was proposed and its exact NSM was computed in \cite{agrell24}.

In dimensions $13$ and $14$, the obtained lattices have lower NSMs than the previously best reported values \cite{lyu22}. These new lattices are parametric and cannot be fully identified by the technique in Sec.~\ref{s:exact}. The $13$-dimensional lattice is composed of eight copies of $A_7\times D_5\times \Z$ and the $14$-dimensional one of four copies of $K'_{10}\times D_4$, with the constituent sublattices scaled separately. These lattices, including their analysis methods and optimal scale factors, are described in \cite{pook-kolb24arxiv}.

\begin{table*} \centering
\caption{Conjectured optimal lattice quantizers obtained by numerical optimization.}
\label{t:results}
\begin{tabular}{ccccc}
\hline
$n$ & $\hat{G} \pm 2 \hat\sigma$ & $\theta(q)$ & Converged lattice & $G$ \\
\hline\hline
$10$ & $0.070811\pm 0.000003$ &
  $1+180q^2+512q^{5/2}+3380q^4+5120q^{9/2}+\cdots$ &
  $D^+_{10}$ & $0.070813818$ \\
$11$ & $0.070424\pm 0.000002$ &
  $1+132q^{3\alpha}+990q^{4\alpha} +2970q^{6\alpha}+12408q^{7\alpha}+\cdots$ &
  $A_{11}^3$ & $0.070426259$ \\
$12$ & $0.070029\pm 0.000002$ &
  $1+120q^2+768q^{5/2}+1024q^3+4104q^4+\cdots$ &
  Glued $D_6\times D_6$ & $0.070031226$ \\
$13$ & $0.069696\pm 0.000002$ &
  $1+56q^{1.989}+40q^{2.006}+2q^{2.049}+256q^{2.412}+\cdots$ &
  Glued $A_7\times D_5\times \Z$ & $0.069697639$\\
$14$ & $0.069261\pm 0.000002$ &
  $1+24q^{1.996}+270q^{2.311}+1920q^{2.731}+960q^{3.467}+\cdots$ &
  Glued $K'_{10}\times D_4$ & $0.069261779$\\
$15$ & $0.068869\pm 0.000002$ &
  $1+280 q^{12\beta}+2048q^{15\beta}+1710q^{16\beta}+5376q^{20\beta}+\ldots$ &
  $\Lambda^*_{15}$ &  $0.068871726$ \\
$16$ & $0.068296\pm 0.000002$ &
  $1+4320 q^{2\sqrt{2}}+61440q^{3\sqrt{2}}+522720q^{4\sqrt{2}}+\cdots$ &
  $\Lambda_{16}$ & $0.068297622$ \\
\hline
&& where $\alpha = 2^{9/11}\cdot 3^{-10/11}$ and $\beta=2^{-12/5}$ &&
\end{tabular}
\end{table*}

In dimension $15$, the algorithm unexpectedly converged to $\Lambda^*_{15}$, the dual of the laminated lattice $\Lambda_{15}$ defined in \cite{conway82laminated}. Among four inequivalent laminated lattices denoted by $\Lambda_{15}$ in \cite{plesken93}, the relevant one is labelled $15.1$. As far as we know, neither $\Lambda_{15}$ nor its dual has been previously considered for quantization. A previously unpublished generator matrix for $\Lambda^*_{15}$ is
\eqlab{e:lx15}{
\renewcommand\arraystretch{1.0}  %
\bB =
\left[\begin{array}{c@{\;\;}c@{\;\;}c@{\;\;}c@{\;\;}c@{\;\;}c@{\;\;}c@{\;\;}c@{\;\;}c@{\;\;}c@{\;\;}c@{\;\;}c@{\;\;}c@{\;\;}c@{\;\;}c}
2 & 0 & 0 & 0 & 0 & 0 & 0 & 0 & 0 & 0 & 0 & 0 & 0 & 0 & 0 \\
0 & 2 & 0 & 0 & 0 & 0 & 0 & 0 & 0 & 0 & 0 & 0 & 0 & 0 & 0 \\
1 & 1 & 1 & 0 & 0 & 0 & 0 & 0 & 0 & 0 & 0 & 0 & 0 & 0 & 0 \\
0 & 0 & 0 & 2 & 0 & 0 & 0 & 0 & 0 & 0 & 0 & 0 & 0 & 0 & 0 \\
0 & 0 & 1 & 1 & 1 & 0 & 0 & 0 & 0 & 0 & 0 & 0 & 0 & 0 & 0 \\
0 & 1 & 0 & 1 & 0 & 1 & 0 & 0 & 0 & 0 & 0 & 0 & 0 & 0 & 0 \\
1 & 0 & 0 & 1 & 0 & 0 & 1 & 0 & 0 & 0 & 0 & 0 & 0 & 0 & 0 \\
0 & 0 & 0 & 0 & 0 & 0 & 0 & 2 & 0 & 0 & 0 & 0 & 0 & 0 & 0 \\
0 & 0 & 1 & 0 & 0 & 0 & 0 & 1 & 1 & 0 & 0 & 0 & 0 & 0 & 0 \\
0 & 1 & 0 & 0 & 0 & 0 & 0 & 1 & 0 & 1 & 0 & 0 & 0 & 0 & 0 \\
1 & 0 & 0 & 0 & 0 & 0 & 0 & 1 & 0 & 0 & 1 & 0 & 0 & 0 & 0 \\
1 & 0 & 0 & 1 & 0 & 0 & 0 & 1 & 0 & 0 & 0 & 1 & 0 & 0 & 0 \\
0 & 1 & 0 & 1 & 0 & 0 & 0 & 1 & 0 & 0 & 0 & 0 & 1 & 0 & 0 \\
0 & 0 & 1 & 1 & 0 & 0 & 0 & 1 & 0 & 0 & 0 & 0 & 0 & 1 & 0 \\
\h & \h & \h & \h & \h & \h & \h & \h & \h & \h & \h & \h & \h & \h & \h     
\end{array}\right]
.}
We computed its NSM numerically and exactly. A numerical estimate of the NSM using the method described in Sec.~\ref{s:estimation} but with a higher accuracy than in Tab.~\ref{t:results} ($10^{12}$ random samples) is $\hat{G}\pm 2\hat\sigma = 0.06887171 \pm 0.00000002$.
The Voronoi region of $\Lambda^*_{15}$ was completely charted using the methods in \cite{pook-kolb22, pook-kolb23}. This computation was more demanding than expected, because the Voronoi region turned out to have $17\,675\,520\,000$ vertices.
For comparison, the best $16$-dimensional lattice has only $201\,343\,200$ vertices \cite{pook-kolb23}.
The kissing number of $\Lambda^*_{15}$ is $280$. Its Voronoi region has $40\,134$ facets and altogether $142\,450\,921\,496\,795$ faces in all dimensions from $0$ to $15$.
These are split into $3\,784\,046$ equivalence classes under its
symmetry group, whose order of $41\,287\,680$ agrees with the value for $\Lambda_{15}$ given in \cite[Table~1]{plesken93}. 
The numbers of classes in dimensions $0$ through $15$ are, respectively,
     $820$, %
 $13\,701$, %
 $92\,610$, %
$330\,483$, %
$703\,579$, %
$956\,395$, %
$862\,632$, %
$526\,405$, %
$219\,298$, %
 $62\,987$, %
 $12\,823$, %
  $1\,994$, %
     $272$, %
      $40$, %
       $6$, %
and    $1$. %
The exact NSM of $\Lambda^*_{15}$ is
\eq{
G &= \frac{306\,171\,250\,566\,121\,481\,924\,149\,919}{3\,369\,080\,934\,926\,681\,018\,204\,160\,000 \cdot 2^{2/5}} \notag\\
    &\approx 0.0688717257963967
,}
which matches the numerical estimate very well. This is a clear improvement over the previously best known lattice quantizer, whose NSM is approximately 0.07037 \cite{lyu22}.

In dimension $16$, finally, the optimization process again converged to the previously best known lattice quantizer. The $16$-dimensional Barnes--Wall lattice, which is also equivalent to the laminated lattice $\Lambda_{16}$, was proposed for quantization already in \cite{conway84} and its exact NSM was computed in \cite{pook-kolb23}.

\section{Conclusions}

Using a new numerical optimization algorithm and a new technique to derive exact lattices from numerically defined ones, conjectured optimized lattice quantizers are obtained in selected dimensions up to $16$. In dimensions $13$ and $14$, new parametric lattices are discovered. The dual of the $15$-dimensional laminated lattice emerges as a new best lattice in dimension $15$. In dimensions $10$, $11$, $12$, and $16$, numerical evidence is obtained that the previously best known lattices might indeed be optimal.

It took $27$ years until the benchmark algorithm was improved and the range of numerically designed lattice quantizers was extended from $10$ dimensions up to $16$. If anyone in the next decades would embark on a journey towards dimensions beyond $16$, we believe that the algorithms reviewed in Sec.~\ref{s:algorithms} may deserve to be carefully explored and refined.

\section*{Acknowledgment}

The authors are grateful to Or Ordentlich for insightful feedback on an earlier version of the manuscript and for suggesting and deriving the alternative optimization algorithms in Sec.~\ref{s:fullmatrix} and \ref{s:whitening}.

\balance

\end{document}